\documentclass[prl,twocolumn,superscriptaddress]{revtex4-1}
\usepackage{epsfig}
\usepackage{graphicx}
\usepackage{palatino}
\usepackage[english]{babel}
\usepackage{hyphenat}
\usepackage{amsmath}
\usepackage{amssymb}
\usepackage{mathtools}
\usepackage{mathrsfs}
\usepackage{slashed}
\usepackage{epstopdf}
\usepackage{xcolor}
\usepackage{booktabs}
\definecolor{lcolor}{rgb}{0.,0.0,0.}
\definecolor{citcolor}{rgb}{0,0.,0.5}
\usepackage[breaklinks,colorlinks,urlcolor=blue,citecolor=blue,linkcolor=blue]{hyperref}
\usepackage{multirow}
\usepackage{ltablex}
\usepackage{soul}

\newcommand{\beq}{\begin{eqnarray}}
\newcommand{\eeq}{\end{eqnarray}}

\newcommand{\bem}{\begin{multline}}
\newcommand{\eem}{\end{multline}}
\newcommand{\beg}{\begin{gather}}
\newcommand{\eeg}{\end{gather}}

\newcommand{\nn}{\nonumber\\}

\newcommand{\ben}{\begin{eqnarray*}}
\newcommand{\een}{\end{eqnarray*}}

\newcommand{\eqn}[1]{Eq.~\eqref{#1}}

\newcommand{\secn}[1]{Section~1}
\newcommand{\appn}[1]{Appendix~1}

\long\def\comment#1{ }

\def\and{\quad\text{and}\quad}

\newcommand{\dif}{{\rm d}}
\newcommand{\rmd}{{\rm d}}

\newcommand{\rme}{{\rm e}}

\newcommand{\qhat}{\hat{q}}
\newcommand{\vect}[1]{\boldsymbol{#1}_{\perp}}

\newcommand{\kt}{\vect{k}}
\newcommand{\xt}{\vect{x}}
\newcommand{\at}{\vect{a}}

\def\0{{\boldsymbol 0}}

\def\rmd{\textrm{d}}

\newcommand{\abar}{\bar{\alpha}_s}

\begin{document}

\title{Anomalous diffusion in QCD matter }
\author{Paul Caucal} 
\email{pcaucal@bnl.gov}
\affiliation{Physics Department, Brookhaven National Laboratory, Upton, NY 11973, USA}
\author{Yacine Mehtar-Tani} 
\email{mehtartani@bnl.gov}
\affiliation{Physics Department, Brookhaven National Laboratory, Upton, NY 11973, USA}
\affiliation{RIKEN BNL Research Center, Brookhaven National Laboratory, Upton, NY 11973, USA}

\begin{abstract}
We study the effects of quantum corrections on transverse momentum broadening of a fast 
parton passing through dense QCD matter. We show that, at leading logarithmic accuracy the broadening distribution tends at late times or equivalently for large system sizes $L$ to a universal distribution that only depends on a single scaling variable $k^2_\perp/Q^2_s$ where the typical transverse momentum scale increases with time as $\ln Q_s^2 \simeq (1+2 \beta ) \ln L - \frac{3}{2}(1+\beta )\,\ln\ln L$ 
 up to non-universal terms, with an anomalous dimension $\beta  \sim  \sqrt{\alpha_s} $. This property is analogous to geometric scaling of gluon distributions in the saturation regime
 and traveling waves solutions to reaction-diffusion processes.  We note that since $\beta  >0$ the process is super-diffusive, which is also reflected at large transverse momentum where the scaling distribution exhibits a heavy tail $k_\perp^{4-2\beta }$ akin to L\'{e}vy random walks.  
\end{abstract}

\maketitle

Transverse momentum broadening (TMB) of energetic quarks and gluons while traversing QCD matter plays a central role in a variety of processes studied at colliders to probe QCD, ranging from jet suppression in heavy ion collisions \cite{Blaizot:2015lma,Qin:2015srf}
to transverse momentum dependent gluon distribution function that encode information on the 3D  structure of the proton and nuclei in high energy collisions
in particular at small Bjorken $x$ where gluon saturation is expected to take place \cite{Lipatov:1996ts,Gelis:2010nm,Albacete:2014fwa}.

High energy partons experience random kicks in hot or cold nuclear matter that cause their transverse momentum $\kt$ w.r.t. to their direction of motion to increase over time. The leading order elastic process is given by a single gluon exchange via Coulomb scattering and leads to an approximate brownian motion in transverse momentum space, so long as the mean free path is larger than the in-medium correlation length, where the typical transverse momentum square scales linearly with system size $L$, namely $\langle k_\perp^2\rangle_{\rm typ} \sim \hat q \, L $, where $\hat q$ is the diffusion coefficient. 

Moreover, radiative processes can also increase transverse momentum of the leading parton due to recoil effects. This question was recently addressed, and it has been shown that such contributions,
albeit suppressed by the coupling constant $\alpha_s$, are enhanced by double logarithms which must be resummed to all orders when $\alpha_s \ln^2 L \sim 1$ \cite{Liou:2013qya,Blaizot:2013vha,Blaizot:2014bha,Iancu:2014kga}.

In this letter we go beyond this result by investigating in more detail the consequences of the non-local nature of the quantum corrections on the TMB distribution. We find in particular that the latter reaches a universal scaling solution at late times (large $L$) that we compute analytically along with its sub-asymptotic deviations exploiting a formal analogy between the present problem and traveling wave solutions to reaction-diffusion processes \cite{Munier:2003vc}.  As a consequence of the self-similarity characterizing the anomalous random walk, the TMB distribution is of L\'{e}vy type. It is in particular associated with a heavy power law tail describing long rare steps which extends over a large range of transverse momenta above the typical transverse momentum scale. 

L\'{e}vy flights are ubiquitous in nature and span a wide variety of stochastic processes in biological systems \cite{viswanathan1996levy,edwards2007revisiting}, molecular chemistry \cite{ZUMOFEN1994303}, optical lattice \cite{PhysRevLett.79.2221}, turbulent diffusion and polymer transport theory \cite{shlesinger1993strange,1995LNP...450.....S}.  Furthemore, heavy tailed distributions are also observed in self-organized critical states \cite{bak1987self,bak1988self}. In this work, we point out for the first time another occurrence of such random walks in the transport of eikonal partons in dense QCD matter and we compute the anomalous exponents that characterize the deviation to standard diffusion. 
 
\section{Quantum corrections to transverse momentum broadening in QCD media}
\label{sec:DLAresum}

The TMB distribution is related to the forward scattering amplitude $\mathcal{S}(\xt)$ of an effective dipole in color representation $R=A,F$ with transverse size $\xt$(see e.g. \cite{Blaizot:2013vha,Kovchegov:2012mbw,DEramo:2010wup}) via a Fourier transform,
\begin{equation}
 \mathcal{P}(\kt)=\int\dif^2\xt \rme^{-i\kt\cdot\xt}\mathcal{S}(\xt)\,.\label{eq:scatt-ampl}
\end{equation}
Considering the dipole formulation in position space allows for a straightforward resummation of multiple scattering by exponentiating the single scattering cross-section, so long as the interactions between the dipole and the medium are local and instantaneous. Thus, we may write $\mathcal{S}(\xt)=\exp(-\frac{1}{4}\frac{C_R}{N_c}\, \qhat(1/\xt^2)L\, \xt^2)$ \footnote{We use bold type for two dimensional transverse vectors and note $a_T=|\at|$.}. The latter relation defines the quenching parameter in the adjoint representation which is assumed to be a slowly varying function of $\xt$. At tree-level it reads
\begin{equation}
 \qhat^{(0)}(1/\xt^2)=\qhat_0\ln\left(\frac{1}{\xt^2\mu^2}\right)\,,\label{eq:qhat-tree}
\end{equation}
up to powers of $x_T \mu$ suppressed terms. For a weakly coupled QGP the bare quenching parameter $\qhat_0$ and the infrared transverse scale $\mu^2$ \footnote{$\mu^2$ can be obtained from the hard thermal loop value of the collision rate \cite{Aurenche:2002pd}. They read respectively $\qhat_0=\alpha_sN_cm_D^2T$ and $\mu=m_D\rme^{-1+\gamma_E}/2$ \cite{Barata:2020rdn}, with $T$ the plasma temperature, $m_D$ the Debye mass. }
 are related to the Debye screening mass in the QGP or to the inverse nucleon size in a nucleus.
 
It is customary to define the emergent saturation scale $Q_s(L)$ via the relation $\mathcal{S}(\xt^2=1/Q_s^2(L))\equiv \mathrm{e}^{-1/4}$, or equivalently, $\qhat(Q_s^2(L))L\equiv Q_s^2(L)$. This definition is standard in small-$x$ physics \cite{Kowalski:2003hm,Lappi:2011ju}, and is also motivated by Moli\`ere theory of multiple scattering \cite{Moliere+1948+78+97,PhysRev.89.1256,Barata:2020rdn} in which $Q_s$ is the transverse scale that controls 
the transition between the multiple soft scattering and the single hard scattering regimes. Given Eq.\,\eqref{eq:qhat-tree}, one finds $Q_s^2\sim \qhat_0L\ln(\qhat_0L/\mu^2)$ at tree level.

Beyond leading order in $\alpha_s$, one has to account for real and virtual gluon fluctuations in the effective dipole with lifetime $\tau$ smaller than the system size. Such fluctuations yield potentially large contributions of the form $\qhat^{(1)}\sim\alpha_s\qhat^{(0)}\ln^2(L/\tau_0)$ with $\tau_0 \ll L  $ a microscopic scale of order of the mean free path \cite{Liou:2013qya}. These radiative corrections to the quenching parameter can be resummed to double logarithmic accuracy (DLA) via an evolution equation ordered in 
$\tau$~\cite{Liou:2013qya,Blaizot:2014bha,Iancu:2014kga}: 
\begin{align}
\qhat(\tau,\kt^2)&=\qhat^{(0)}+\int_{\tau_0}^{\tau}\frac{\dif\tau'}{\tau'}\int_{Q^2_{ s}(\tau')}^{\kt^2}\frac{\dif \kt'^2}{\kt'^2} \ \abar \ \qhat(\tau',\kt'^2)\,,\label{eq:qhat-DL}\\
Q^2_{s}(\tau)&=\qhat(\tau,Q_{ s}^2(\tau))\tau\,,\label{eq:Qsat}
\end{align}
where $\abar=\alpha_sN_c/\pi$ and $\qhat^{(0)}\equiv \hat q (\tau_0,\kt)$ .
The condition $\kt^2\ge Q^2_s(\tau)$ in Eq.\,\eqref{eq:qhat-DL} enforces the gluon fluctuations to be triggered by a single scattering with plasma constituents whose contribution is logarithmically enhanced  compared to multiple scattering. 
The final value of $\tau$ is fixed at the largest time allowed by the saturation condition, i.e.\ at the time scale $\tau$ such that $Q_s^2(\tau)=\kt^2$, as long as $\tau<L$ and $\tau =L$ otherwise. The latter case corresponds to the dilute regime \cite{Blaizot:2019muz}, that is when $\kt^2 \gg Q_s^2(L)$.

In this letter, we address both analytically and numerically  the non-linear system \eqref{eq:qhat-DL}-\eqref{eq:Qsat} \footnote{See Supplemental Material 1 for detailed explanations about the numerical resolution of this non-linear differential system.}. Analytic solutions are in general difficult to obtain, however, a solution for the linearized problem that consists in approximating $Q_s^2(\tau)\simeq \qhat_0\tau$ for the emission phase space can be found \cite{Iancu:2014sha,Mueller:2016xoc}. For a constant initial condition
it reads 
 \begin{align}
  \qhat(Y,\rho)&=\qhat_0\left[\textrm{I}_0\left(2\sqrt{\abar Y\rho}\right)-\frac{Y}{\rho}\textrm{I}_2\left(2\sqrt{\abar Y\rho}\right)\right]\,,\label{eq:analytic-sol}
 \end{align}
 with $Y=\ln(\tau/\tau_0)$ and $\rho=\ln(\kt^2/(\qhat_0\tau_0))$ \footnote{We note $\mathrm{I}_n$ the modified Bessel function of rank $n$}. Formally, this linearization is valid in DLA, which captures all the terms of the form $\abar^n Y^{2n}$ assuming $Y \sim \rho$, but it misses sub-leading corrections of the form $\abar^nY^{2n-1}\ln Y $ which are parametrically larger than the single logarithmic ones.
This is one of the novelty of the present study, enabling us to highlight the geometric scaling property of the transverse momentum diffusion in QCD and to compute the scaling deviations.

\section{Geometric scaling and traveling waves}
\label{sec:geoscal}

The TMB distribution is said to obey geometric scaling if it is only a function of $\kt^2/Q^2_s(L)$ as a result of scaling invariance of the radiative process at late times. More precisely, we would have 
\begin{equation}
 \lim\limits_{L\to\infty}\qhat(\kt^2)L=Q_s^2(L)f\left(\frac{\kt^2}{Q_s^2(L)}\right)\,,
\end{equation}
for some scaling function $f$ to be determined. Note that the linearized version of Eq.\,\eqref{eq:qhat-DL} satisfies a similar scaling relation with the notable difference that the argument of $f$ is $\kt^2/(\qhat_0L)$ instead of $\kt^2/Q_s^2(L)$. The non-linearity of Eq.\,\eqref{eq:qhat-DL} enforces the evolution to be controlled by a single momentum scale $Q_s(L)$ \footnote{Note that this property holds also for the tree level form of $\qhat$ as $\qhat^{(0)}(\kt^2)L\simeq Q_s^2(L)$ for $k_T\ll Q_s^2(L)/\mu$.}.

Geometric scaling was extensively studied in the context of deep inelastic scattering, where it has been shown that the gluon distribution $g(x,Q^2)$ at small $x$ satisfies this property over a broad region of photon virtuality $-Q^2$ \cite{Stasto:2000er,Iancu:2002tr,Kwiecinski:2002ep}. 
We shall demonstrate that TMB exhibits similar properties.

\paragraph{Traveling waves (TW) solution.} Remarkably, it is possible to find the scaling function $f$ for the non-linear problem defined by Eqs.\,\eqref{eq:qhat-DL}-\eqref{eq:Qsat}. In terms of the variables $Y$ and $\rho$, the differential equation satisfied by $\qhat$ reads
\begin{equation}
  \frac{\partial \qhat(Y,\rho)}{\partial Y}=\abar \int_{\rho_{s}(Y)}^{\rho}\rmd\rho' \ \qhat(Y,\rho')\,,\label{eq:eqndiff_Yrho}
\end{equation}
with $\rho_s=\ln(Q_s^2/(\qhat_0\tau_0))$.
Let us now look for a scaling solution of the form $\qhat(Y,\rho)=\qhat_0e^{\rho_s(Y)-Y}f(\rho-\rho_s(Y))$. 

Plugging this expression in \eqn{eq:eqndiff_Yrho}, one gets the following differential equation
\begin{equation}
 -\frac{\rmd \rho_s}{\rmd Y}f''(x)+\left[\frac{\rmd \rho_s}{\rmd Y}-1\right]f'(x)-\abar f(x)=0\label{eq:eqndiff-f}\,,
\end{equation}
where $x= \rho-\rho_s(Y)$. In order for $f$ to be a function of $x$ only at large $Y$, the derivative $\rmd \rho_s/\rmd Y$ must converge towards a constant $c$, which can be interpreted as the speed of a traveling wave that propagates to the right on the $\rho$ axis. This is reminiscent of the TW solutions \cite{Munier:2003vc} to the Balitsky-Kovchekov (BK) equation \cite{Balitsky:1995ub, Kovchegov:1999yj}. The initial conditions to this second order linear differential equation follow from the non-linear treatment of the saturation boundary: on the isoline $x=0$, meaning $\rho=\rho_s(Y)$, one has $f(0)=0$ and $f'(0)=(c-1)/c$. It is then straightforward to solve \eqn{eq:eqndiff-f} with the scaling form $f(x)\propto  \rme^{\beta x }$ where the slope $\beta$ is solution of the quadratic equation $ -c \beta^2 +(c-1) \beta -\abar =0$. For any initial condition such that $\qhat(Y=0,\rho) < \rme^{\beta  \rho}$ when $\rho\to \infty$ \cite{Brunet:1997zz}, which is the case in the present letter,  the minimum value for $\beta$, that satisfies the additional constraint $-2 c \beta  +(c-1)=0$, will be realized. This fixes the speed to $c=1+2\sqrt{\abar+\abar^2}+2\abar\simeq 1+2\sqrt{\abar}$.  Hence, for this critical value that minimizes the TW speed, the solution to \eqref{eq:eqndiff-f} takes the form 
\begin{equation}\label{eq:scale-sol}
 f(x)=\left(1+\beta  x\right) \rme^{\beta  x}\quad, \quad \beta \equiv\frac{c-1}{2c}\,.
\end{equation}

In terms of the physical variables the $\kt^2$ dependence of $\qhat$ that enters the broadening distribution reads, in the large $L$ limit, 
\begin{align}
 \frac{\hat q(\kt^2)L}{Q_s^2(L)}=\begin{cases}
              \left(\frac{\kt^2}{Q_s^2(L)}\right) ^{2\beta } & \hspace{-2cm}\textrm{if } \kt^2\le Q_s^2(L)\label{eq:qhat-geoscal}\\
                  \left(\frac{\kt^2}{Q_s^2(L)}\right) ^{\beta } \left[1+\beta \ln\left(\frac{\kt^2}{Q_s^2(L)}\right)\right] & \\ \qquad \qquad  \qquad\qquad\textrm{otherwise}\,,\\                
               \end{cases}
\end{align}
which is continuous and derivable everywhere. 

\begin{figure}[t]
 \centerline{\includegraphics[width=0.95\columnwidth]{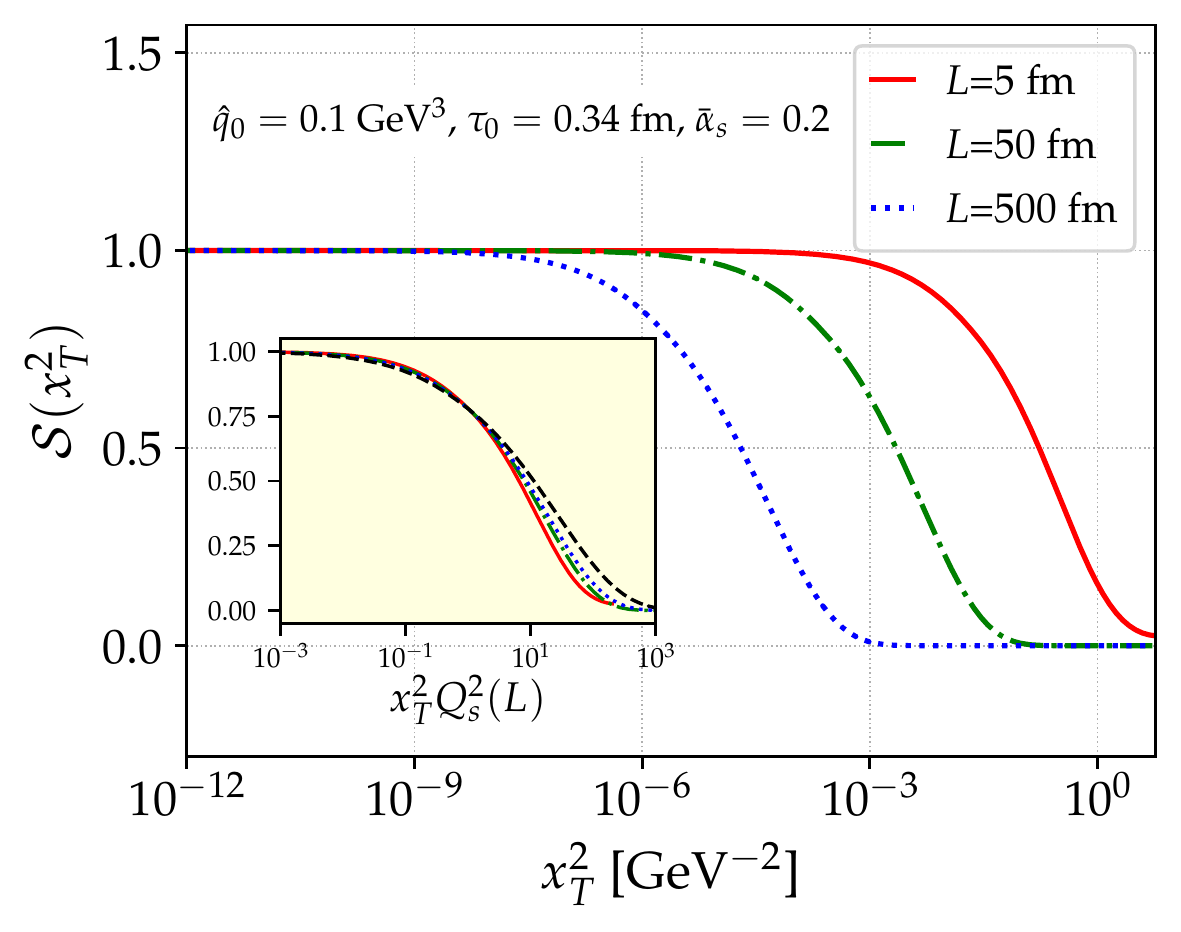}}
 \caption{ Dipole scattering amplitude as a function of the transverse dipole size squared $x_T^2$ for several values of the medium size. The inset shows the same curves as a function of the scaling variable $x_T^2Q_s^2(L)$ compared to the scaling solution (dashed black) (color online). 
 }\label{fig:pt-broad-1}
\end{figure}

To make the interpretation of these results in terms of TW's more transparent we inserted Eq.\,\eqref{eq:qhat-geoscal} in $\mathcal{S}(\xt)$ and plotted the result in Fig.\,\ref{fig:pt-broad-1} for several values of $L$. We see that the traveling wave propagates from right to left (from large to small $x_T$) with increasing $L$. However, once plotted in terms of $x_T^2Q_s^2$ as shown in the inset of Fig.\,\ref{fig:pt-broad-1}, they all lie approximately on the same universal scaling function given by Eq.\,\eqref{eq:scatt-ampl}-\eqref{eq:qhat-geoscal} (dotted black curve). The observed deviations will be discussed in what follows. 

\paragraph{Sub-asymptotic corrections.} We turn now to the calculation of the sub-asymptotic corrections to the geometric scaling solution \eqref{eq:qhat-geoscal}. 
Near the wave front, typically for $x\gg 1$, we can  look for a solution of the form
\begin{align}
 \qhat(Y,\rho)&=\qhat_0\rme^{\rho_s(Y)-Y} \rme^{\beta x} Y^\alpha G\left(\frac{x}{Y^\alpha}\right)\,, \\
 \rho_s(Y)&=cY+b\ln(Y)\,,\label{eq:LsY-subleading}
\end{align}
inspired by the traveling waves ansatz that solves the BK equation \cite{Munier:2003sj} and more generally FKPP-like equations \cite{fisher1937,10003528013}. Plugging this ansatz inside Eq.\,\eqref{eq:eqndiff_Yrho}, one gets a differential equation for $G(z)$.
Because the coefficient of $Y$ and $Y^\alpha$ (assuming $\alpha>0$) in this equation must vanish, we recover the two previous constraints that fix the value of $c$ and $\beta$. 
Then, neglecting the power suppressed terms $Y^{-1}$ and $Y^{-\alpha-1}$, one finds
\begin{align}
 &-c Y^{-\alpha}\frac{\rmd^2}{\rmd z^2}G(z)-\beta \alpha z Y^{\alpha-1}\frac{\rmd }{\rmd z}G(z)\nn
 &+\beta  (-b\beta +b+\alpha)Y^{\alpha-1}G(z)=0\,.
\end{align}
The homogeneity condition implies that the coefficient $\alpha$ must be equal to $1/2$ so that the deviation from the scaling form near the wave-front grows in a diffusive way as $Y$ increases.

For initial conditions such that $\beta< \beta $, the large $z$ behavior of the function $G(z)$ constrains the acceptable values of the coefficient in front of $G(z)$ in this differential equation. In fact, as shown in \cite{Brunet:1997zz,Munier:2003sj}, one must have $b\beta -b-\frac{1}{2}=1$, fixing $b$ to 
\begin{align}
 b&=-\frac{3}{2(1-\beta )}\,.
\end{align}
This yields the solution 
\begin{equation}
 G\left(z=x/\sqrt{Y}\right)=\beta  z\exp\left(-\frac{\beta  z^2}{4c}\right)\,.\label{eq:Gsol}
\end{equation}
The value of $b$ we extract from this analysis is novel and a consequence of the non-linearity of the saturation boundary. In the linearized problem with the lower bound in the integral of Eq.\,\eqref{eq:eqndiff_Yrho} set to $Y$ instead of $\rho_s(Y)$, one gets $b=-3/2$ from the analytic solution \eqref{eq:analytic-sol} \cite{Iancu:2014sha}, whereas in the non-linear case, we have $b\simeq -3/2(1+\sqrt{\abar})$. The sub-leading term provides a correction to the saturation line which is parametrically of order $\sqrt{\abar}\ln Y$ and therefore dominant w.r.t\, single logarithmic corrections of order $\abar Y \sim \sqrt{\abar} $ since in DLA $Y \sim 1/\sqrt{\abar} \gg 1$. 

The TW solution \eqref{eq:Gsol} provides the functional form of $\qhat(Y,\rho)$ near the wave front, i.e.\ for $x=\rho-\rho_s(Y)\sim \sqrt{Y}\gg1$ and fixes the value of the coefficient $b$ in the asymptotic expansion of $\rho_s(Y)$. For small values of $x$, one can find the scaling deviations by looking for a solution as a power series in $1/Y$ of the form $\qhat(Y,\rho)=\qhat_0e^{\rho_s(Y)-Y}f(x)\sum_{n\ge0}\frac{H_n(x)}{Y^n}$. Plugging this form inside Eq.\,\eqref{eq:eqndiff_Yrho} gives a second order differential equation for $H_1(x)$, whose initial conditions are constrained by the definition of $Q_s$. The solution to this equation reads:
$H_1(x)=b x / [c^2(1+\beta  x)] [1+(c-1)(c+3)/ (8c) x+ (c-1)^2(1+c)/(48c^2)x^2]$ \footnote{See Supplemental Material 2 for detailed calculations.}. 
The last term in this expression is included in the solution \eqref{eq:Gsol}, as can be checked by expanding the function $G$ for large $Y$, but not the first two since Eq.\,\eqref{eq:Gsol} is only valid at large $x$. Combining the scaling limit with its deviation provided by the function $G$ for $x\sim \sqrt{Y}$ and $H_1(x)$ for all $x$ up to powers of $1/Y^2$, our final result reads 
\begin{align}
& \qhat(Y,\rho)=\qhat_0\, \rme^{\rho_s(Y)-Y}\exp\left(\beta  x-\frac{\beta  x^2}{4cY}\right)\nonumber\\
 &\times\left[1+\beta  x+\frac{bx}{c^2Y}\left(1+\frac{\beta (c+4)x}{6}\right)+\mathcal{O}\left(\frac{1}{Y^2}\right)\right]\,,\label{eq:qhat-final}
\end{align}
with $\rho_s(Y)=cY+b\ln(Y)+\rm const$. 

This solution is independent of the initial condition (for physically relevant ones), and only depends on the value of $\abar$ via the coefficients $c$, $\beta $ and $b$.  
The resummed TMB distribution displays a universal behavior independent of the non-perturbative modeling of the tree-level distribution often used as an initial condition for non-linear small $x$ evolution \cite{McLerran:1993ni,McLerran:1993ka}. It can therefore provide a model-independent functional form for the initial condition of the BK equation, that includes gluon fluctuations enhanced by double logs, $\abar \ln^2 A^{1/3}$, inside the nucleus target to all orders.  
 
\begin{figure}[t]
 \centerline{\includegraphics[width=0.95\columnwidth]{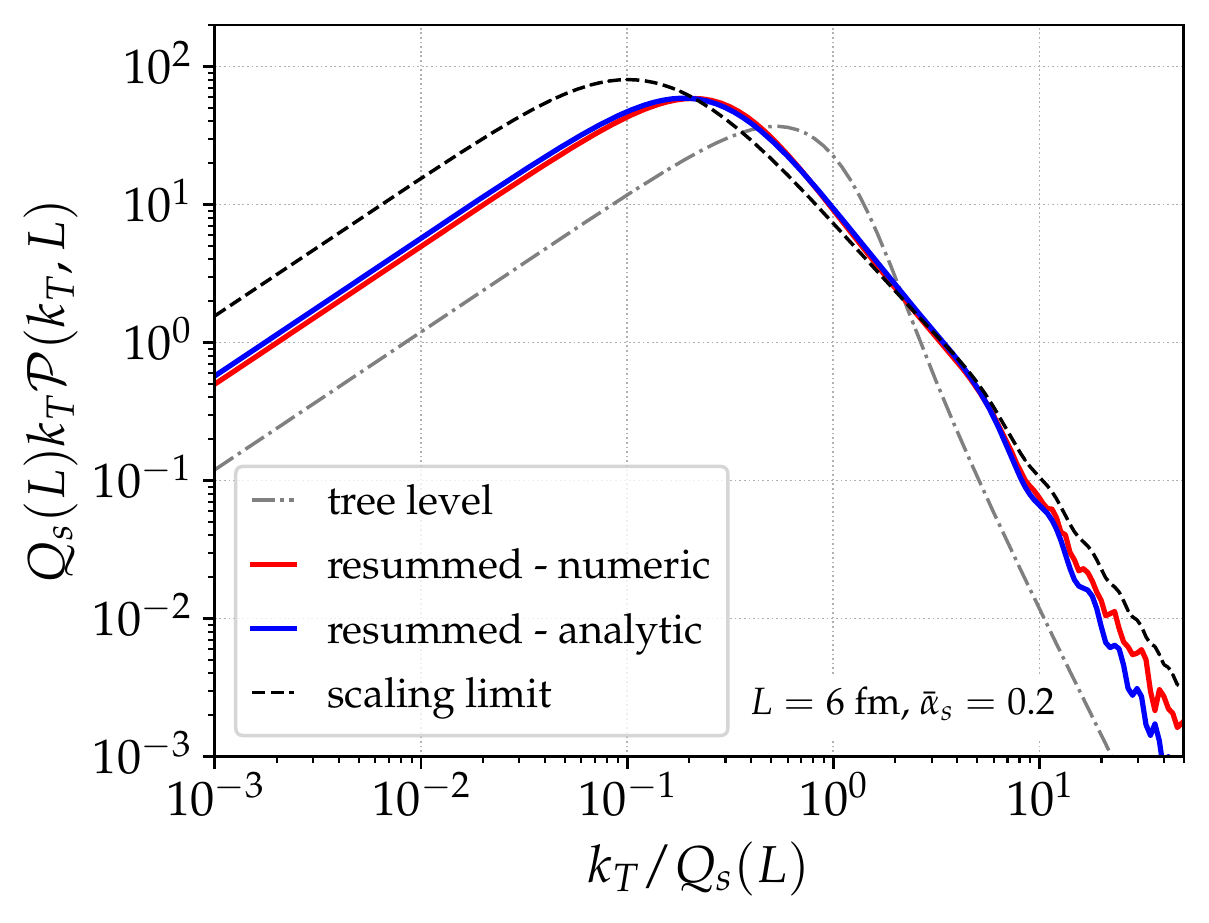}}
 \caption{TMB distribution of a high energy gluon propagating though a dense medium of size $L$ at tree-level (dash-dotted line) and after resummation of the leading radiative corrections (solid red).The dotted black line is the scaling limit when $L\to\infty$ and the blue curve is our analytic result given by Eq.\,\eqref{eq:qhat-final} including sub-asymptotic corrections (color online). 
}
  \label{fig:pt-broad-2}
\end{figure}

%
\section{Super-diffusion and modification of Rutherford scattering}
\label{sec:supdiff}

In this last section, we investigate the physical consequences of the scaling solution \eqref{eq:qhat-geoscal} for $\qhat(\kt^2)$ on the TMB distribution given by Eq.\,\eqref{eq:scatt-ampl}, in particular at large $k_T$, where the distribution is characterized by rare events that are sensitive to the point-like nature of the medium scattering centers \cite{DEramo:2012uzl}. First, it is straightforward to see that in the large $L$ limit, the TMB distribution $\mathcal{P}(k_T)$ is only a function of $k_T/Q_s(L)$.
In Fig.\,\ref{fig:pt-broad-2}, we show the distribution as a function of $k_T/Q_s(L)$ with $Y=\ln(L/\tau_0)=4$, for the following set-ups: (i) in dash-dotted grey, at tree-level, using Eq.\,\eqref{eq:qhat-tree}, (ii) after quantum evolution obtain by numerically solving Eqs.\,\eqref{eq:Qsat} in red, (iii) in blue, using the expression Eq.\,\eqref{eq:qhat-final} that includes sub-asymptotic corrections to the scaling limit, (iv) finally, in dashed black, the scaling limit $Y\to\infty$ of Eq.\,\eqref{eq:qhat-final}. Interestingly, the sub-asymptotic corrections account for the relatively large deviations between the asymptotic curve and the exact numerical result at the moderate value of $L=6$ fm.

The $k_T$ distribution exhibits two different regimes: the region of the peak, near $Q_s(L)$ and the large $k_T$ tail, with $k_T\gg Q_s(L)$. These results can be interpreted in term of a special kind of random walk (here in momentum space) called L\'{e}vy flight. Such a remarkable connection with statistical physics enables us to highlight some interesting features  (i) self-similar dynamics (ii) super diffusion (iii) power-law tail with slower decay than the Rutherford $\kt^{-4}$ behavior. 

In order to further the connection with the physics of anomalous diffusion, consider the scaling limit of the TMB distribution in the vicinity of the peak where the shape of the distribution is controlled by the first line in Eq.\,\eqref{eq:qhat-geoscal}. Using this solution, one  finds that $\mathcal{S}(\xt) \simeq \exp\left[ - \,\frac{1}{4}\frac{C_R}{N_c}\, ( |\xt|Q_s)^{2-4\beta }\right]$.  In momentum space, it implies that the distribution $\mathcal{P}(\kt,L)$ satisfies a generalized Fokker-Planck equation,
$ \partial \mathcal{P}/\partial L\propto\ -(-\Delta)^{1-2\beta }\mathcal{P}$, 
where the so-called fractional Laplace operator $ (-\Delta)^{\gamma/2}$ is defined by its Fourier transform $|\xt|^\gamma$ \cite{intro-frac-deriv,kwasnicki2017ten}. This fractional diffusion equation (without external potential)  is satisfied by the probability density for the position of a particle undergoing a L\'{e}vy flight process in two dimension \cite{dubkov2008levy} with stability index $\gamma=2-4\beta \simeq2-4\sqrt{\alpha_s}+\mathcal{O}(\alpha_s)$.

\begin{figure}[t]
\centerline{ \includegraphics[width=0.95\columnwidth]{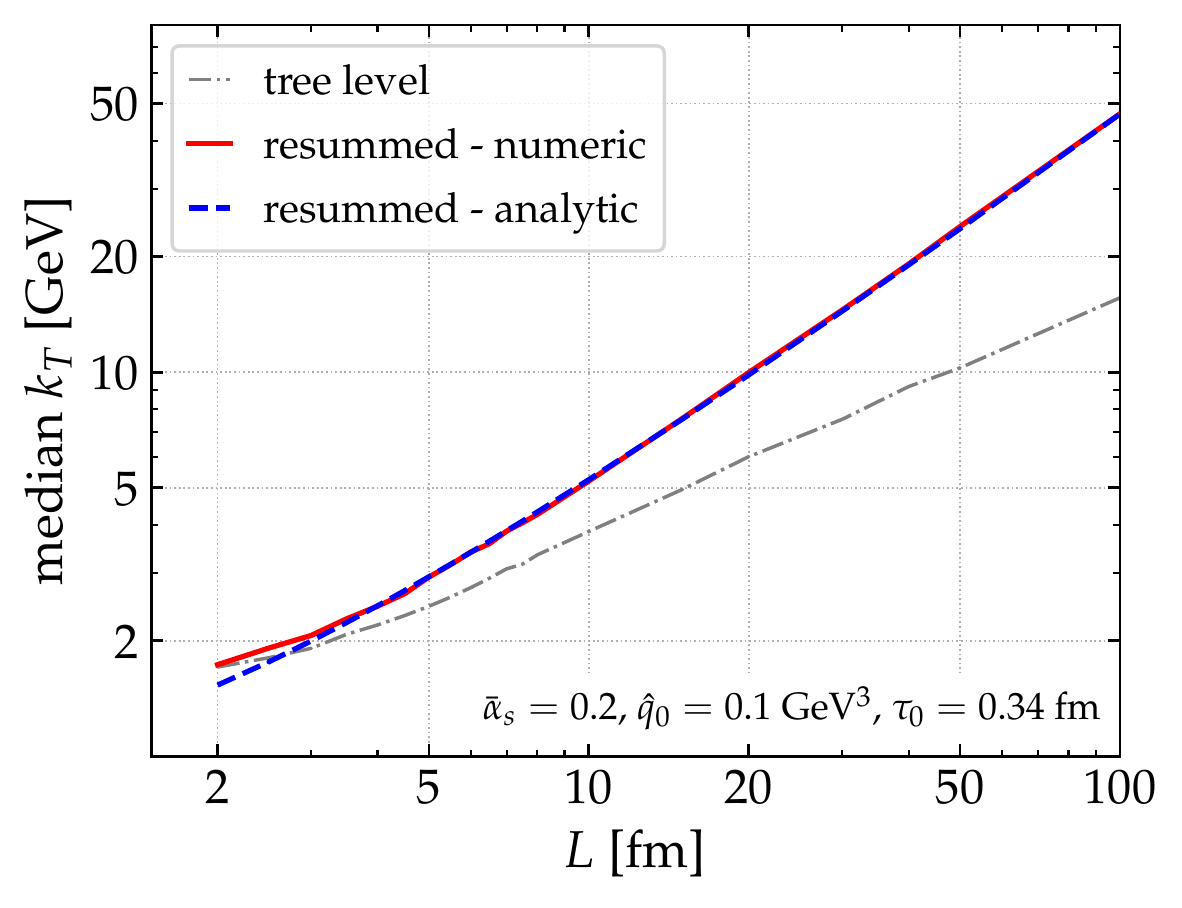}}
 \caption{System size dependence of the median of the TMB distribution at tree-level (grey line) and after numerical resummation of radiative corrections (red line). The dashed blue line is our analytic prediction given by Eq.\,\eqref{eq:LsY-subleading}.}
  \label{fig:median-kt}
\end{figure}

Because of its heavy tail  (to be discussed thereafter), the mean $k_T^2$ of the TMB distribution is not defined. Nevertheless, it is possible to introduce a measure of the characteristic width of the $k_T$ distribution, and study its behavior as a function of the medium size $L$. In what follows, we shall use the median value $\langle k_T\rangle_{\rm med}$ of $k_T\mathcal{P}(k_T)$ \footnote{ Another possibility is to define fractional moments \cite{metzler2002space,dubkov2008levy}} which is shown in Fig.~\ref{fig:median-kt} for three different scenarios. In grey, we plot the tree-level resulting from Eq.~\eqref{eq:scatt-ampl} and \eqref{eq:qhat-tree}. The median scales approximately like $(L \ln L)^{1/2}$, which up to the logarithmic factor resulting from the Coulomb logarithm in \eqn{eq:qhat-tree},  exhibits the standard diffusion scaling.
 The red line is the median of the $k_T$ distribution obtained using the resummed value of $\qhat$ with fixed coupling, after numerical resolution of Eqs.\,\eqref{eq:Qsat}. We then compare this result with our analytic prediction \eqref{eq:LsY-subleading} (assuming $\langle k_T\rangle_{\rm med}\propto Q_s$ \footnote{the unknown pre-factor is determined by a fit to the numerical result at large $L$}), $\langle k_T\rangle_{\rm med}\propto L^{^{\frac{c}{2}+\frac{b}{2}\frac{\ln(Y)}{Y}}}$,
which is represented in blue in Fig.\,\ref{fig:median-kt} .
Remarkably, the agreement is excellent down to rather small values of $L\sim 3$ fm.
Since $c/2>1/2$, the median grows faster than $\sqrt{L}$ at large $L$, illustrating the super-diffusive behavior of TMB beyond leading order, with a deviation to the standard diffusion of order $\sqrt{\abar}$.

\paragraph{Heavy-tailed distribution.} Another important aspect of L\'{e}vy flights is
the power law decay of the step length distribution for a L\'{e}vy walker \cite{klafter1987stochastic,metzler1999deriving}. This reflects the fact that long jumps with arbitrary length may occur with non-negligible probability. In the problem at hand, this power-law tail can also be understood as a consequence of the self-similar nature of overlapping successive gluon fluctuations.

The tail of the TMB distribution is controlled by the large $\kt^2$ behavior of $\qhat(\kt^2)$, and consequently, by the exponential in the second line in \eqref{eq:qhat-geoscal}.  
Note, however, that the scaling limit encompasses two stability indices: one controlling the peak and the median, as discussed above, while the other controls the tail of the distribution (cf. \eqref{eq:qhat-geoscal}). Without loss of generality, one can derive the leading behavior of $\mathcal{P}(\kt)$ at large $\kt$ by expanding the dipole S-matrix for small dipole sizes, as a result the Fourier transform can be approximated by \footnote{See Supplemental Material 3 for detailed calculations.}:
\begin{align}
 \mathcal{P}(\kt) \sim \vec{\nabla}_{\kt}^2\frac{\pi}{\kt^2}\frac{\dif \qhat (\kt^2)L }{\dif \ln \kt^2 }\,,\label{eq:P1-largekt}
\end{align}
up to logarithmically suppressed terms. 
This formula quantifies the deviations from the Rutherford scattering cross-section that are induced by radiative corrections.
Applying Eq.~\eqref{eq:P1-largekt} to our scaling solution \eqref{eq:qhat-geoscal}, one finds the tail
 \begin{equation}
  \mathcal{P}(\kt)\propto\frac{1}{Q_s^2(L)}\left(\frac{Q_s^2(L)}{\kt^2}\right)^{\nu}\,,
 \end{equation}
 with $\nu=2-\beta +\mathcal{O}(\ln(x)/x)$ and $x=\ln(\kt^2/Q_s^2(L))$. The corrections to the power-law behavior are due to the pre-factor in the second line of \eqref{eq:qhat-geoscal}.
 The power of the tail deviates from the tree level Rutherford behavior by $\sim -2\sqrt{\abar}$. The form of $\nu$ is correct in the strict scaling limit $L\to \infty$. For finite $L$ values, the $1/\kt^4$ tail is recovered at very large $k_T$, as can be inferred from Eq.\,\eqref{eq:analytic-sol} which yields  $\nu=2-2\sqrt{\abar Y/x}$ (when $x\gg Y$). The fact that geometric scaling extends in the tail region is known in the context of saturation physics as the ``extended geometric scaling window" corresponding to $Q_s\ll k_T \ll Q_s^2/\mu$.  
 
Finally, note that this analysis is valid so long as $E/k_T^2 \gg L$, where $E$ is the energy of the fast parton. This allows us to neglect the quantum diffusion of the dipole in the medium. In the opposite case the quantum phase is suppressed when $E/k_T^2  \ll \tau \ll L$ and the evolution is expected to be of DGLAP type with the substitution $\ln L/\tau_0 \to \ln  E/(k_T^2\tau_0)$ \footnote{This regime was discussed in \cite{Casalderrey-Solana:2007xns}.}.

\paragraph{Conclusion.} In summary, we have studied the transverse momentum distribution of a high-energy parton propagating through a dense QCD medium, including resummation of radiative corrections within a modified double logarithmic approximation which accounts for 
the non-linear dynamics due to multiple scatterings that restrict the phase space for quantum fluctuations. We have found that 
the non-linearity and self-similarity of overlapping multiple gluon radiations lead to a universal scaling limit at large system sizes, which exhibits a super-diffusive regime and a power-law decay akin to L\'{e}vy flights. Although at very high $\kt$, the distribution is characterized by point-like interactions of Rutherford type, for moderately large $\kt$ we observe a weaker power due to the non-local nature of the interactions which is the hallmark of scale invariant phenomena. 

Concerning phenomenological applications, we point out the relevance of our analytic solutions in the study of nuclear structure at high energy as it provides a new initial condition for non-linear evolution of the gluon distribution. We leave for future work the question of the experimental detection of this emergent QCD phenomenon in heavy ion collisions as well as running coupling corrections which are expected to yield mild scaling violations.

\smallskip
\noindent{\bf Acknowledgements}
 This work was supported by the U.S. Department of Energy, Office of Science, Office of Nuclear Physics, under contract No. DE- SC0012704. Y. M.-T. acknowledges support from the RHIC Physics Fellow Program of the RIKEN BNL Research Center.
 
\bibliographystyle{apsrev4-1}
\bibliography{biblio}
\clearpage
\appendix

\begin{widetext}
\section{1. Numerical implementation of the evolution equation}
\label{app:A}

\paragraph{Differential problem.} We discuss here the numerical method used to solve the evolution equation for $\qhat$ with exact treatment of the saturation boundary for the gluon emission phase space. The evolution equation in differential form reads
\begin{align}
 \frac{\partial \qhat(\tau,\kt^2)}{\partial\ln( \tau)}&=\abar \int_{Q^2_{s}(\tau)}^{\kt^2}\frac{\dif  \kt'^2}{\kt'^2}\qhat(\tau,\kt'^2)\\
 Q^2_{s}(\tau)&=\qhat(\tau,Q_{s}^2(\tau))\tau\,.\label{eq:Qsat-ap}
\end{align}
The existence of a solution is not guaranteed
since the boundary of the $\kt'$ integral depends on $\qhat(\tau,\kt^2)$ itself. That said, it is possible to show that the two equations above constitute a well defined initial-value problem. To do so, we differentiate the l.h.s. of Eq.~\eqref{eq:Qsat-ap} w.r.t. $\tau$:
\begin{align}
 \frac{\dif Q_{s}^2(\tau)}{\dif \tau}&=\frac{\dif \qhat(\tau,Q_{s}^2(\tau))}{\dif \tau}\tau+\qhat(\tau,Q_{s}^2(\tau))\,,\\
 &=\frac{\partial\qhat(\tau,\kt^2)}{\partial\kt^2}\Bigg|_{\kt^2=Q^2_{s}(\tau)}\frac{\dif Q^2_{s}(\tau)}{\dif \tau}\tau+\frac{1}{\tau}Q_{s}^2(\tau)\,.
\end{align}
The partial derivative of $\qhat(\tau,\kt^2)$ can be obtained from the evolution equation in its integral form:
\begin{align}
 \qhat(\tau,\kt^2)&=\qhat(\tau_0,\kt^2)+\abar\int_{\tau_0}^{\tau}\frac{\dif \tau'}{\tau'}\int_{Q^2_{s}(\tau')}^{\kt^2}\frac{\dif \kt'^2}{\kt'^2}\qhat(\tau',\kt'^2)\,,\\
 &=\qhat(\tau_0,\kt^2)+\abar\int_{Q^2_{s}(\tau_0)}^{\kt^2}\frac{\dif \kt'^2}{\kt'^2}\int_{\tau_0}^{\mathrm{min}(\tau,\tau_{s}(\kt'^2))}\frac{\dif \tau'}{\tau'}\qhat(\tau',\kt'^2)\,,
\end{align}
with $\tau_s(\kt^2)$ the inverse function of $Q^2_s(\tau)$ that satisfies then $\tau_s(Q^2_s(\tau))=\tau$. One can then compute the derivative w.r.t. $\kt^2$, evaluated at $Q_s^2(\tau)$:
\begin{align}
 \frac{\partial\qhat(\tau,\kt^2)}{\partial\kt^2}\Bigg|_{\kt^2=Q^2_{s}(\tau)}=\frac{\partial\qhat(\tau_0,\kt^2)}{\partial\kt^2}\Bigg|_{\kt^2=Q^2_{s}(\tau)}+\frac{\abar}{Q^2_{s}(\tau)}\int_{\tau_0}^{\tau}\frac{\dif \tau'}{\tau'}\qhat(\tau',Q^2_{s}(\tau))\,.
\end{align}
Finally, we end up with the following differential equation for $Q^2_{s}(\tau)$:
\begin{equation}
 \left(1-\tau\frac{\partial\qhat(\tau_0,\kt^2)}{\partial\kt^2}\Bigg|_{\kt^2=Q^2_{s}(\tau)}-\frac{\abar\tau}{Q^2_{s}(\tau)}\int_{\tau_0}^{\tau}\frac{\dif \tau'}{\tau'}\qhat(\tau',Q^2_{s}(\tau))\right)\frac{\dif Q_{s}^2}{\dif  \ln(\tau)}=Q^2_{s}(\tau)\,.
\end{equation}
This equation is more convenient from a numerical point of view because one does not need to solve the implicit equation defining $Q^2_{s}(\tau)$ at each small step in $\tau$. Instead, one solves two coupled integro-differential equations of order 1.

The initial condition $\qhat(\tau_0,\kt^2)$ enables to define a \textit{bare} saturation boundary. For example, we may assume that $\qhat(\tau_0,\kt^2)=\qhat_0f_0(\kt^2)$ with some function $f_0(\kt^2)$.
Defining $Y=\ln(\tau/\tau_0)$ and $\rho = \ln\left(\frac{\kt^2}{\qhat_0\tau_0}\right)$, and the following dimensionless quantities 
\begin{align}
 \qhat_d(Y,\rho)=\frac{\qhat\left(\tau_0\rme^{Y},\qhat_0\tau_0\rme^\rho\right)}{\qhat_0}\,,\qquad \rho_{s}(Y)=\ln\left(\frac{Q^2_{s}(\tau_0\rme^{Y})}{\qhat_0\tau_0}\right)\,,
\end{align}
the system of differential equation reads finally
\begin{align}
  \frac{\partial \qhat_d(Y,\rho)}{\partial Y}&=\abar \int_{\rho_{s}(Y)}^{\rho}\dif \rho'\qhat_d(Y,\rho'\,,)\label{eq:diffqs}\\
\frac{\dif \rho_{s}(Y)}{\dif Y}&=1+ \rme^{Y-\rho_{s}(Y)}\left[\left.\frac{\dif  f_0}{\dif \rho}\right|_{\rho=\rho_{s}(Y)}+\abar\int_{0}^{Y}\dif  Y'\qhat_d(Y',\rho_{s}(Y))\right]\frac{\dif \rho_{s}(Y)}{\dif Y}\,.\label{eq:diffLsat}
\end{align}
The second equation deserves some comments: (i) the first ($1$)  term is to be associated with the linearized saturation line $Q^2_{s}(\tau)=\qhat_0\tau$, since the solution of $\frac{d\rho_{s}(Y)}{dY}=1$ with $\rho_{s}(0)=0$ is $\rho_{s}(Y)=Y$ which gives $Q^2_{s}(\tau)=\qhat_0\tau$, 
(ii) the term proportional to $\alpha_s$ inside the square brackets  is the back-reaction of the quantum evolution of $\qhat$ on the saturation boundary.

\paragraph{Euler method.} This evolution equation is solved using the Euler method, by discretizing the time $Y\in[0,Y_{L}=\ln(L/\tau_0)]$. The transverse momentum space is also discretized $\rho\in[0,\rho_{\rm max}]$ and the $\rho$ integrals are evaluated using the trapezoidal rule. The initial condition $\qhat_d(0,\rho)=f_0(\rho)$ to the differential problem is given by the tree-level functional form of $\qhat_d(\rho)$, in which we keep the leading logarithmic dependence.
In the hard thermal loop calculation of $\qhat$ at leading order, this logarithmic dependence reads $\qhat(\kt^2)=\qhat_0\ln(\kt^2/\mu^2)$ (cf.\ Eq.~\eqref{eq:qhat-tree})
with $\qhat_0=\frac{g^2}{4\pi}C_Rm_D^2T$ and $\mu^{2}=\frac{m_D^2}{4}\rme^{-2+2\gamma_E}$.
Using this expression as the initial condition $\qhat(\tau_0,\kt^2)$ at the initial time $\tau_0$, the defining equation for $Q^2_{s}(\tau_0)$ reads $Q_{s}^2(\tau_0)=\qhat(Q_{s}^2)\tau_0$ or equivalently $\exp\left(\rho_{s}(0)\right)=\rho_{s}(0)+\ln\left(\frac{\qhat_0\tau_0}{\mu^{2}}\right)$.
This equation has a real solution if and only if
\begin{equation}
\ln\left(\frac{\qhat_0\tau_0}{\mu^{2}}\right)\geqslant 1\Longrightarrow\tau_0\geqslant\frac{\rme\mu^{2}}{\qhat_0}=\frac{\rme^{-1+2\gamma_E}\pi}{g^2T}\,.
\end{equation}
Thus, we shall use $\tau_0=\rme^{-1+2\gamma_E}\pi/(g^2T)$ which is of the order of the mean free path $\sim 1/g^2T$. With this choice, $\rho_{s}(0)=0$ and 
\begin{equation}
 f_0(\rho)=\rho+1,\,\qquad \frac{\dif f_0(\rho)}{\dif\rho}=1\,.
\end{equation}
From a numerical standpoint, the initial condition $f_0(\rho)=\rho+1$ requires some care because the naive Euler method fails at $Y=0$, since from Eq.~\eqref{eq:diffLsat}, the derivative of $\rho_s(Y)$ w.r.t. $Y$ in $Y=0$ diverges. One needs then to initialize the Euler method in the middle of the first bin using
\begin{equation}
 \rho_{s}\left(\frac{\Delta Y}{2}\right)=\frac{\Delta Y}{2}+\ln\left[-\mathrm{W}\left(-\rme^{-\frac{\Delta Y}{2}-1},-1\right)\right]\,,
\end{equation}
where $\mathrm{W}(x,n)$ is the Lambert function on the $n^{\rm th}$ branch.

\section{2. Sub-asymptotic corrections away from the wave front}
\label{app:B}

In this section, we compute the sub-asymptotic corrections to the scaling solution, in the regime where $x$ is not necessarily large. We try the following ansatz:
\begin{align}
 \qhat_d(Y,\rho)&=e^{\rho_s(Y)-Y}f(x)\left(1+\frac{bH_1(x)}{Y}+{\cal O}\left(\frac{1}{Y}\right)\right)\,,\label{eq:scaling-violation1}\\
 \rho_s(Y)&=c Y+b\ln(Y)+{\cal O}(\ln(Y))\,.\label{eq:scaling-violation2}
\end{align}
The form is inspired by the large $Y$ expansion of the travelling wave ansatz. We have indeed, with $G$ given by Eq.\,\eqref{eq:Gsol}:
\begin{equation}
 \qhat_d(Y,\rho)\simeq e^{\rho_s(Y)-Y}\beta x e^{\beta x}\left(1-\frac{\beta x^2}{4cY}+{\cal O}\left(\frac{1}{Y}\right)\right)\,,
\end{equation}
for $x\gg1$ and large $Y$.
Now, we insert Eqs.\,\eqref{eq:scaling-violation1}-\eqref{eq:scaling-violation2} inside the differential equation \eqref{eq:diffqs} satisfied by $\qhat_d$, and we treat $1/Y$ as a small parameter in the large $Y$ limit. Note that $1/Y$ terms are generated by the derivative of $\rho_s$ with respect to $Y$. The zeroth order term gives back the differential equation satisfied by the scaling function $f$,
\begin{equation}
 (c-1)f(x)-cf'(x)=\abar\int_0^x\dif x'\,f(x')\,,
\end{equation}
with solution $f(x)=e^{\beta  x}(1+\beta  x)$ when $c=1+2\sqrt{\abar+\abar^2}+2\abar$, with $\beta =(c-1)/(2c)$. On the other hand the $\mathcal{O}(1/Y)$ term gives
\begin{equation}
 f(x)+(c-1)H_1(x)f(x)-cf(x)H_1'(x)-cH_1(x)f'(x)-f'(x)=\abar\int_0^x\dif x'\,H_1(x')f(x')\,.
\end{equation}
We thus need to solve this equation to determine $H_1(x)$. Remarkably, defining $u(x)=H_1(x)f(x)$, we end up with the simpler differential equation
\begin{align}
 (c-1)u(x)-cu'(x)-\abar\int_0^x\dif x'\,u(x')&=f'(x)-f(x)\,.
\end{align}
Differentiating with respect to $x$, this equation reduces to:
\begin{align}
 (c-1)u'(x)-cu''(x)-\abar u(x)&=f''(x)-f'(x)\,,
\end{align}
with initial condition $u(0)=0$, $u'(0)=1/c^2$. These initial conditions comes from the definition of the saturation boundary $\qhat(\tau,Q_s^2(\tau))\tau=Q_s^2(\tau)$.
Given that $c=1+2\sqrt{\abar+\abar^2}+2\abar$, we solve this differential equation for $u(x)$, leading to the following expression for $H_1(x)$:
\begin{equation}
H_1(x)=\frac{x}{c^2(1+\beta  x)}\left[1+\frac{(c-1)(c+3)}{8c}x+\frac{(c-1)^2(1+c)}{48c^2}x^2\right]\,.
\end{equation}

\section{3. Transverse momentum broadening in the large $\kt$ regime}
 
In this section we study without loss of generality the large $\kt$ behavior of Eq.\,\eqref{eq:scatt-ampl}. We note $F(\kt^2)=\qhat(\kt^2)L$.
Expanding the exponential, one gets
\beq\label{eq:P-UV}
\mathcal{P}(\kt) &\simeq & -  \frac{1}{4}  \int \dif^2\xt \,  F(1/\xt^2)\,\xt^2\,   \,  \rme^{-i\xt\cdot \kt }\,,\\
&= & \frac{1}{4} \vec{\nabla}^2_{\kt}   \int \dif^2\xt \,  F(1/\xt^2)\,  \,  \rme^{-i\xt\cdot \kt }\,,\\
&=& \frac{\pi}{2} \vec{\nabla}^2_{\kt}   \frac{1}{\kt^2}\int_0^{+\infty } \dif z\,z  \,   F(1/z^2 \kt^2)\,  \,  \mathrm{J}_0(z)\,.
\eeq
In the limit $k_T \to + \infty$, the above integral is dominated by $z \sim 1 $, and we thus have  $\ln 1/z^2 \ll \ln k_\perp^2$. This provides us with a small expansion parameter $\lambda \equiv \ln\frac{1}{z^2}$ such that 
\beq
 F(1/z^2 \kt^2) &=&  \sum_{n=0}^\infty   \frac{\dif^ n F (\kt^2) }{\dif( \ln \kt^2 )^n}\lambda^n\,.
\eeq
We have therefore the asymptotic series representation of the broadening distribution at large $k_T$,
\beq
 \mathcal{P}(\kt) &\simeq &\frac{\pi}{2}\sum_{n=0}^\infty  \, a_n \vec{\nabla}^2_{\kt} \frac{1}{\kt^2} \,  \frac{\dif^ n F (\kt^2) }{\dif( \ln \kt^2 )^n}\,,\label{eq:Pkt-asymptotic} \eeq
 where  
 \beq
 a_n =  \int_0^{+\infty } \dif z\,z  \,   \ln^n\frac{1}{z^2}\,  \,  \mathrm{J}_0(z)  \,.
 \eeq
 These coefficients can be computed analytically. To do so, one first needs to compute the Mellin transform of the Bessel function $\mathrm{J}_0(z)$,
\begin{equation}
  \int_0^\infty \dif z\, z^j\mathrm{J}_0(z)=2^j\frac{\Gamma\left(\frac{j+1}{2}\right)}{\Gamma\left(\frac{1-j}{2}\right)}\,,
\end{equation}
where $\Gamma$ is the standard gamma function.
Note that for every odd $j$, the integral vanishes. In particular, one finds that the 0-th order in Eq.\,\eqref{eq:Pkt-asymptotic} does not contribute to \eqn{eq:P-UV} since $a_0=0$.
The coefficients $a_n$ can be obtained by Mellin transform using the trick
\begin{equation}
 \ln^n\left(\frac{1}{z^2}\right)=\lim\limits_{\alpha\to0}\frac{\partial^n}{\partial\alpha^n}\left(\frac{1}{z^2}\right)^\alpha\,,
\end{equation}
so that
\begin{equation}
a_n\equiv\int_0^\infty \dif z\, z \ln^n\left(\frac{1}{z^2}\right)J_0(z)=\lim\limits_{\alpha\to0}\frac{\partial^n}{\partial\alpha^n}\,\left[2^{1-2\alpha}\frac{\Gamma\left(1-\alpha\right)}{\Gamma\left(\alpha\right)}\right]\,,
\end{equation}
which gives for the first few terms 
\begin{align}
 a_1&=2\\
 a_2&=4(2\gamma_E-\ln(2))\\
 a_3&=24(\gamma_E-\ln(2))^2,...
\end{align}
Numerically, they read: $a_1=2$, $a_2=-0.93$, $a_3=0.32$, $a_4=38.34$, etc. Truncating the asymptotic series in Eq.\,\eqref{eq:Pkt-asymptotic} up to the first term leads to Eq.\,\eqref{eq:P1-largekt} in the main text.
\end{widetext}

\end{document}